\documentclass[journal]{IEEEtran}
\usepackage{url}
\usepackage{graphicx}
\usepackage{biblatex}
\usepackage{float}
\usepackage{hyperref}
\usepackage{amsmath,amssymb,amsfonts}
\usepackage[linesnumbered,ruled,vlined,onelanguage]{algorithm2e}
\usepackage{graphicx}
\usepackage[dvipsnames]{xcolor}
\usepackage{tikz}
\usepackage{forest}
\usetikzlibrary{backgrounds}
% ...
\tikzset{every picture/.style={show background rectangle}}
\usetikzlibrary{shapes.geometric}
\usetikzlibrary{positioning}
\begin{document}
\title{Development of Bayesian Component Failure Models in E1 HEMP Grid Analysis}

\author{
\IEEEauthorblockN{Niladri Das},
\IEEEauthorblockN{Ross Guttromson},\and
\IEEEauthorblockN{Tommie A. Catanach}\\
\IEEEauthorblockA{\textit{Sandia National Laboratories}, USA}
}
\maketitle
\footnotetext{Sandia National Laboratories is a multimission laboratory managed and operated by National Technology and Engineering Solutions of Sandia, LLC, a wholly owned subsidiary of Honeywell International Inc., for the U.S. Department of Energy’s National Nuclear Security Administration under contract DE-NA0003525.}

\begin{abstract}
Combined electric power system and High-Altitude Electromagnetic Pulse (HEMP) models are being developed to determine the effect of a HEMP on the US power grid.  The work relies primarily on deterministic methods; however, it is computationally untenable to evaluate the E1 HEMP response of large numbers of grid components distributed across a large interconnection.  Further, the deterministic assessment of these components' failures are largely unachievable.  E1 HEMP laboratory testing of the components is accomplished, but is expensive, leaving few data points to construct failure models of grid components exposed to E1 HEMP.  The use of Bayesian priors, developed using the subject matter expertise, combined with the minimal test data in a Bayesian inference process, provides the basis for the development of more robust and cost-effective statistical component failure models. These can be used with minimal computational burden in a simulation environment such as sampling of Cumulative Distribution Functions (CDFs).
\end{abstract}

\section{Introduction}

\begin{figure}[htbp]
\includegraphics[width=\linewidth]{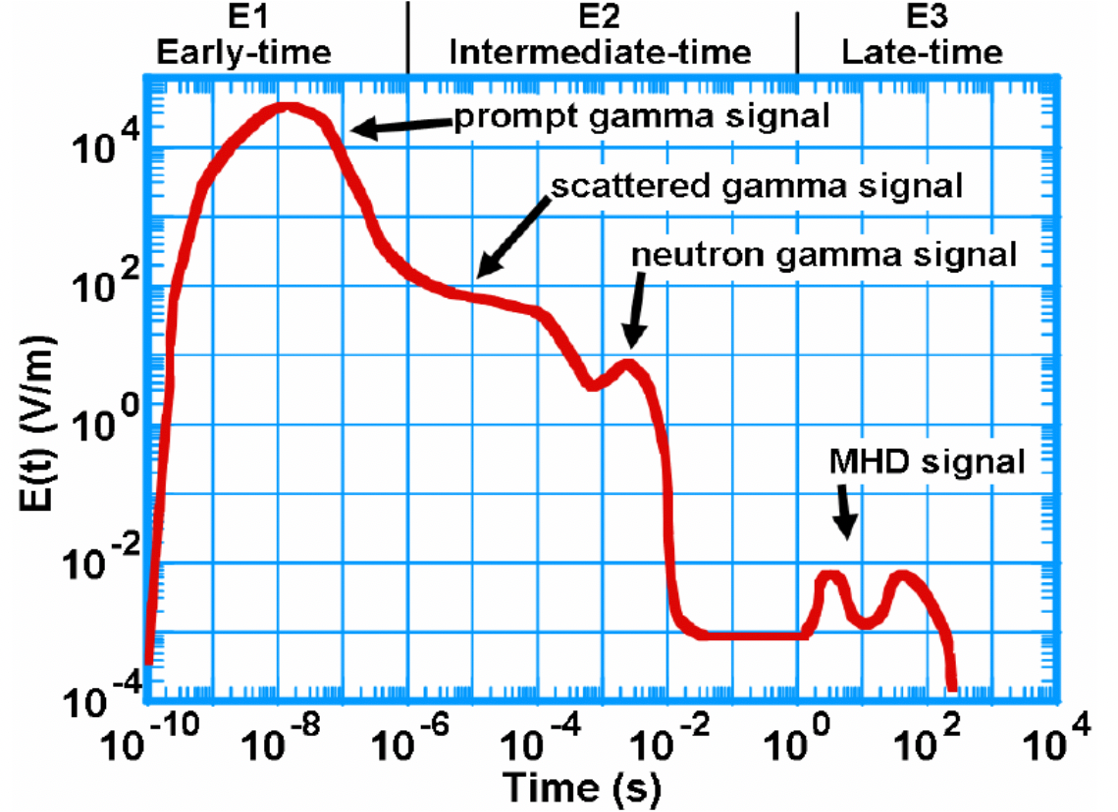}
\caption{High Altitude EMP Waveform \cite{1}}
\label{figHEMP}
\end{figure}

A HEMP event occurs when a nuclear weapon detonates high in the atmosphere; nominally between 75 km and 300 km above ground.  A HEMP is a single electromagnetic pulse and is often categorized into three parts: early-time (E1), intermediate-time (E2), and late-time (E3), as shown in Figure \ref{figHEMP} \cite{1,8}. The three parts result from different physics interactions between the nuclear blast and the upper atmosphere and have different electromagnetic and frequency characteristics, resulting in different types of insults to components and potential damage.

A software simulation tool, the HEMP Transmission Consequence Model (HTCM), is being developed by a team from Sandia National Laboratories, Los Alamos National Laboratories, Texas A\&M and the Western Area Power Administration to evaluate the consequences to a large power grid interconnection from a high-altitude electromagnetic pulse (HEMP), from both E1 HEMP (early time) and E3 (late time) electromagnetic insults. Not all grid components are vulnerable to a HEMP and those which are generally cannot be modeled in a deterministic manner but rather require statistical models for large-scale assessment. The extent to which the US power grid may be vulnerable to the effects of an electromagnetic disturbance generated from HEMP detonation is not well known and is the fundamental purpose of the HTCM model being developed.  

\begin{figure}[htbp]
\centering
\includegraphics[width=\linewidth]{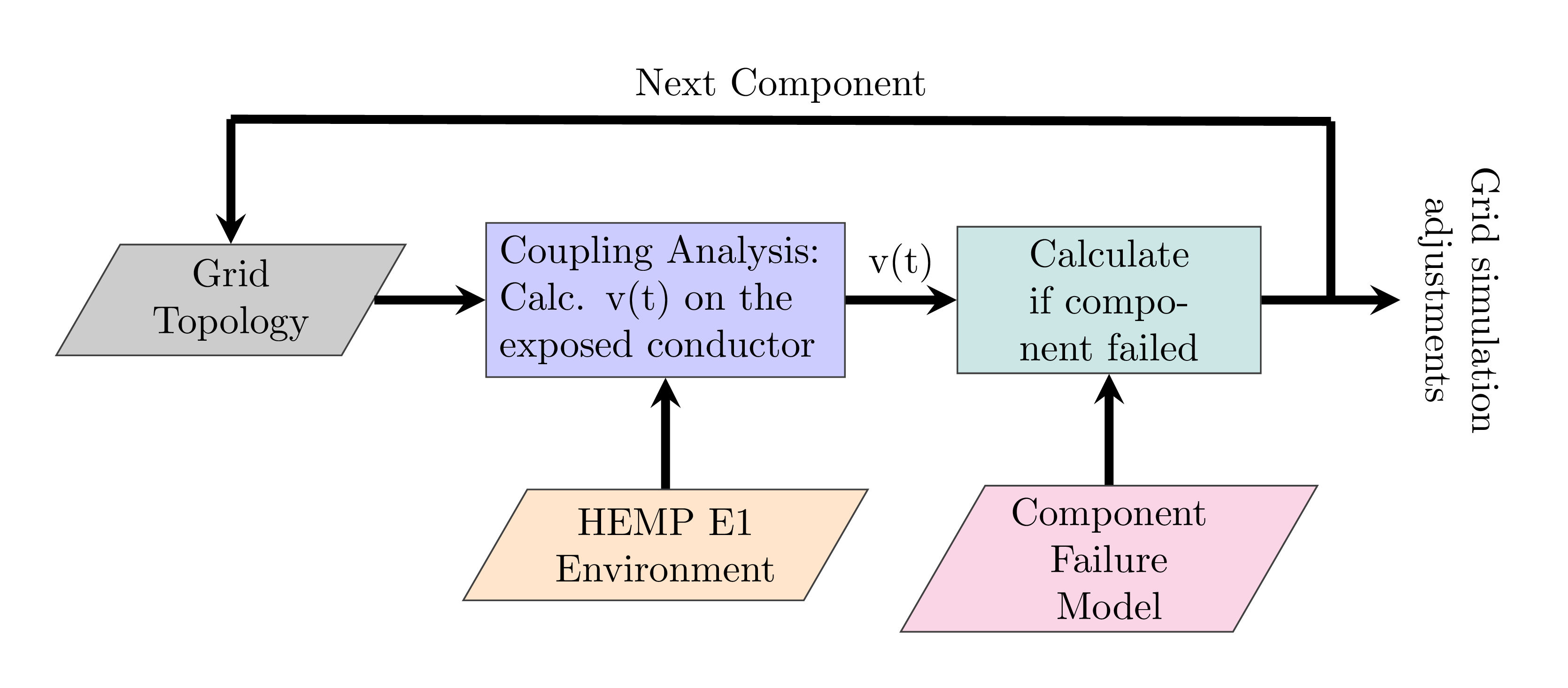}
\caption{Process for Using Failure Model in Grid Simulations \cite{7}}
\label{fig:CouplingScreen}
\end{figure}

Figure \ref{fig:CouplingScreen} shows a basic process diagram of how the HTCM software determines if a HEMP causes a grid component to fail during the simulation.  A Component Failure Model is a statistical failure model in the form of a CDF, conditional upon HEMP coupled voltage, where a failure event is determined by sampling the associated probability density function shown in Eq. \eqref{eqn:CDF}.
Generally, a HEMP can result in the de-energization of electricity to equipment, cause misoperation, loss of functionality, or it can damage equipment directly such that restoration of a power source would not restore the component back to service.  

The methods presented in this paper will focus on the development of statistical component failure models for use in the HTCM model, responding to a simulated E1 HEMP only. This paper will not address E2 HEMP or E3 HEMP.

\section{E1 EMP Component Testing}
It is important to understand the probability of failure of various grid components to a test HEMP pulse in order to understand how they will react to an actual E1 EMP pulse exposure.  Understanding this can inform large-scale grid-HEMP models and aid in the deployment of mitigation technologies where needed.  These tests can be achieved using laboratory-based HEMP testing for each unique type of component as follows.

\subsection{Conducted and Radiated EMP Testing}
Grid components that are potentially susceptible to an E1 HEMP can be tested in a laboratory to determine their susceptibility in one of two ways: exposure to 1) conducted insults and/or 2) radiated insults.  Although combined radiated and conducted insults can occur, few tests have been performed to characterize those effects.  During an actual E1 HEMP, conducted insults occur when a grid component is electrically connected to wires that have been exposed to the HEMP's radiated field, resulting in a coupled voltage and current pulse on the wire \cite{2,5}.  The wires themselves are rarely vulnerable to damage, but act as an antenna, coupling the E1 electromagnetic pulse to the wire and converting it to a high current and high voltage transient pulse.  These pulses travel down the wire and insult connected equipment.  A radiated pulse will directly insult the equipment that is in the E1 HEMP field.  However, this type of insult generally has the greatest effect on electronic equipment \cite{4}. Laboratory testing is one means to determine an EMP's propensity to damage equipment.
In a laboratory environment, a Marx bank can be used to discharge a stored electric charge into a pulse shaping filter which is then connected to a conductive wire or cable (for conducted pulse insults) or an antenna (for radiated pulse insults) which is used to test the component.

\subsection{Test Methods}
The results of laboratory E1 HEMP tests provide a means to statistically characterize test results for use in a large system simulation.  A large number of tests for a given component yield the most accurate results for this method, however results are often determined based on only one or a few test articles since the articles and tests can be expensive \cite{17}, resulting in very low confidence. In most cases, one or two specific brands and models of a component selected for E1 HEMP testing may be used to represent an entire class of components, so their test results do not imply exact results for the class of components, but are used to represent the class.    
The objective of the laboratory E1 HEMP test is to find a probability of failure for the component class, conditional on the magnitude of the peak of the insult voltage, where an increase in voltage reveals a higher probability of failure. Steps for E1 HEMP laboratory testing include:
\begin{enumerate}
  \item Identify important failure modes for the device under test.
  \item Develop a State-of-Health (SOH) detection procedure for these failure modes, caused by the E1 HEMP test insult.
  \item Develop an E1 HEMP insult test procedure and test configuration.
  \item Insult the component, starting at a low insult voltage, gradually increasing between test insults. After each insult check the SOH of the component.
  \item If the component fails its SOH check, note the failure mode and insult voltage. The component is considered to have failed at that insult voltage value.
  \item Using results from the components tested, develop a E1 HEMP statistical component failure model for the component class, that can be used in modeling and simulation.  

\end{enumerate}
\begin{figure}[htbp]
\includegraphics[width=0.45\textwidth]{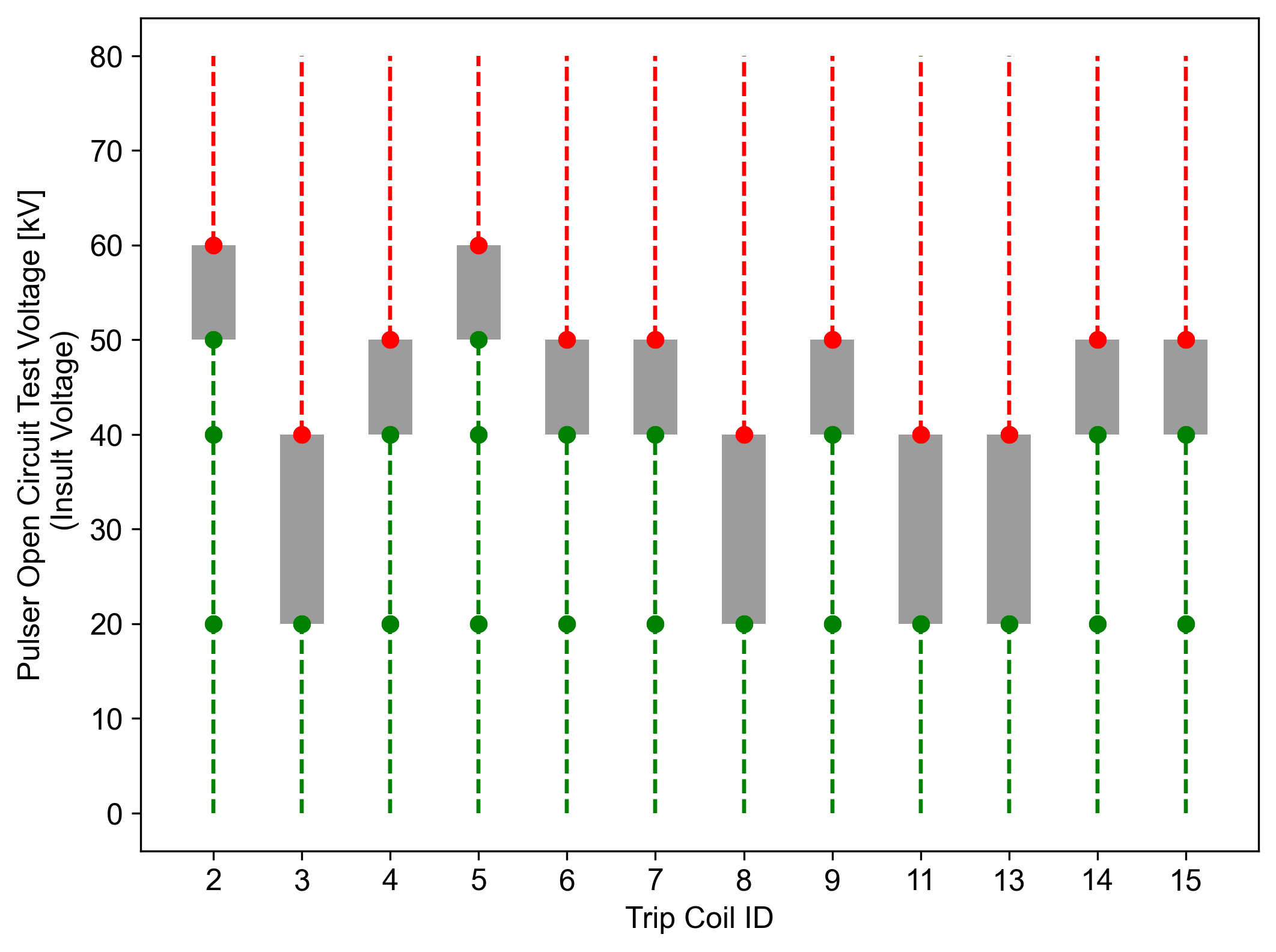}
\caption{Conducted pulse data for 12 breaker trip coils, testing insulation rupture \cite{b15}. Green dots Indicate a \textit{passed} test, while red dots indicate a \textit{failed} test. The gray region is unknown. (Test data for devices 1, 10, and 12 are not available.) }
\label{fig:tripcoildata}
\end{figure}Before the first insult and between each insult, the component's state of health is tested to identify whether the component has retained its full functionality.  Fig. \ref{fig:tripcoildata} shows the pulse test results for 12 identical breaker trip coils. Green dots indicate pulse insults that did not cause failure or damage to the component.  The Marx bank pulser imposes electromagnetic insults typically called shots, which were repeated at increasing voltages until the component failed (owing to component insulation rupture), which occurred well below the pulser's maximum voltage limit   
The vertical gray bars in Fig. \ref{fig:tripcoildata} identify the range in which the failure limit of the component actually exists.

As each component is pulsed, checked, and then repulsed at increasing voltage levels, they may also become weakened or damaged by the test progression, which may influence (decrease) the voltage value at which the component fails.  

Therefore, the authors assumed that the test process caused some degree of dielectric damage during the test sequence. However, avoiding this damage will require testing a new product for each pulse (no retesting) which will be a very costly process to conduct. 

This paper presents an improved method for converting test data into a Bayesian statistical failure model. The first, non-Bayesian method does not consider  dielectric damage to the test article during testing. In the Bayesian method, component damage accumulated during testing is considered as depicted in Fig.\ref{fig:expdataline}.  The values on the x-axis represents the accumulated damage factor which is based on the impulse voltage squared for each test, then normalized using the highest voltage which caused component failure for the set of components tested.  Using the square of the voltage ensures a very minor assumed damage for voltages below the observed failure threshold.  The development of a statistical component failure model that accounts for dielectric damage during its test process (prior to its failure during test) allows for a more realistic failure model-- one which represents the probability of component failure from a HEMP without having been exposed to prior$_{\boldsymbol{\lambda}}$ HEMP insults.  This process is discussed in Section III-B and again in Section IV.

\section{Use and Interpretation of Statistical Component Failure Models in Grid Simulation}
Using a deterministic analysis to identify whether components throughout the grid have failed due to the effects of a HEMP would be intractable and computationally intense. Therefore, a statistical component failure model is used in the power grid simulation to determine which components in the grid model fail, become inoperable, or misoperate due to an E1 HEMP.

In these statistical models, component failure is conditional on the magnitude of the voltage coupled to the cables that are electrically connected to each component. For example, this could include voltage and/or current coupling to a cable that is connected from a metering transformer to a protective relay.  Or a cable connecting a relay output to a breaker trip coil circuit, etc.  However, the statistical failure model itself must be developed using data from the controlled EMP test, as previously described.  Although not presented here, the same process is used to determine statistical failure models for an E1 HEMP \emph{radiated} environment.

Where necessary, a distinction can be made in the types of failure modes associated with a component.  In some cases, the type of failure may result in a different simulation response, such as when a failure results in a component's misoperation as opposed to a complete shutdown of the component. This distinction can be reflected into a statistical failure model by dividing areas under the CDF into failure modes. However, in the work presented here, we consider only one failure mode while developing the statistical component failure model.

There are multiple ways to utilize the test data to generate a statistical failure model. We will show two methods, one using a non-Bayesian technique and a new method using a Bayesian technique that includes SME information into the construction of the statistical failure model, using fewer component tests. This allows for more robust solutions when large test sets are infeasible due to cost or other factors.

\section{Generating a non-Bayesian Statistical Failure Model}

The development of a statistical failure model for a component (or class of component) is typically accomplished in a lab by insulting the component with progressively increasing HEMP pulse magnitudes to determine the level at which it fails.  When `several' components are tested in this way, their results can be combined to establish a probability of failure, conditional upon the insult magnitude.  These data can be represented as a Conditional Cumulative Distribution Function, with varying levels of confidence, dependant upon the number of components tested and their test results.

If $P_{fail}(V_{insult}) $ is the total probability of failure at or below insult voltage  $V_{insult}$ and  $\text{PDF}_{fail}(v)$ is the probability density function (PDF) of failure for a given insult voltage $v$ then:
\begin{align}
    P_{fail}(V_{insult})  = \int_{0}^{V_{insult}} \text{PDF}_{fail}(v)dv.\label{eqn:CDF}
\end{align}

\begin{figure}
    \centering
    \includegraphics[width=1\linewidth]{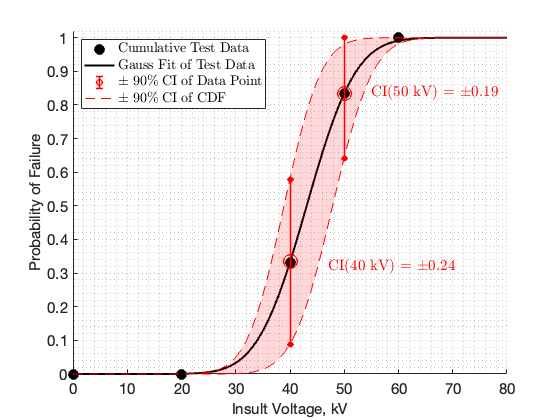}
    \caption{Least Squares Gaussian Fit to Test Data (n=12) with 90 percent Confidence Intervals (CI). Red Dotted Lines Indicate LSE Gaussian Fit of CI with the Cumulative Test Data Having Values 0 and 1. Shape of CDF is Assumed to be Gaussian.}
    \label{fig:enter-label}
\end{figure}

Note that $\text{CDF} \triangleq \int{\text{PDF}}$, so sampling a conditional CDF is represented by Eq. \eqref{eqn:CDF}, which is our definition of a component Statistical Failure Model. The graphical statistical failure model representing the dataset from Fig. \ref{fig:tripcoildata} is shown in Fig. \ref{fig:enter-label}. 

To obtain this statistical failure model, pulse testing is performed to collect data that represents a group of related devices, such as a Class, Brand, or Model.  An example of component test results are shown in Fig. \ref{fig:tripcoildata}, which exhibit test results from identical components, showing insult magnitudes, passes (green) and failures (red). Grey regions are neither pass nor fail.  Although both current and voltage are related to each other by the component input impedance, the voltage has the greater effect on the component's dielectric breakdown.  Subsequent damage is then caused by current.  Hence, our failure probability has been made to be conditional only on the voltage insult level.

The application of Eq. \eqref{eqn:CDF} for a component, using a single test is problematic.  If the test component fails, one must assume a probability of failure of 100\% for similar components subject to the same conditions.  Or if the insult voltage exceeds the pulse voltage capability, this will result in a zero probability of failure for all similar components. Both of these conclusions are incorrect due to several key facts:
\begin{enumerate}
\item Complete certainty about inferring a single test result to a population is not possible.
\item The laboratory test configuration does not represent all field installation configurations.
\item The pulse shape of an actual (future) HEMP event is unknown and may not represent what was applied in laboratory conditions.
\item Variations in manufacturing across vendors and related models may not be consistent with the tested component.
\item The cumulative effects of pulsing until failure may cause various levels of dielectric damage to the component, causing premature failure during the test.
\end{enumerate} 
 Hence, to reliably estimate $\text{P}_{fail}(V_{insult})$ used in Eq. \eqref{eqn:CDF}, a single test result is not sufficient, rather results from multiple component tests are necessary. When there are limited test data available from one class of devices, we may desire to integrate test data from a similarly behaving component class in generating the device failure CDF. The Bayesian method is a well established process to integrate test data into a common statistical result.
This consolidation of multiple Classes of devices with different Brands and Models within it, requires the use of a Bayesian statistical hierarchy model. This is used to develop a statistical component failure model, which learns the model parameters from the test data and generates the component failure CDF.

\section{Development of Bayesian Statistical Failure Model Using Limited Test Data}
 Statistical failure model CDF can be made using non-Bayesian methods if sufficient data exists, but Bayesian methods provide a robust solution when there is minimal test data, but a subject matter expert (SME) failure estimate and/or related test data from related components does exist. With minimal test data (e.g. due to high testing or component costs), the Bayesian approach still offers a method to obtain a robust statistical failure model by fusing test measurements from different sources. This approach can integrate one or more related data sets, limited test data and SME estimates, across classes. The introduction of hierarchy into data sets establishes robustness to the problem of limited test data, and provides a statistically rigorous way to make scientific inferences about the properties of a class of components (or across classes) by using test results from one or many `similar' components and/or an SME failure estimate.

\begin{figure}[htbp]
  \includegraphics[width=0.45\textwidth]{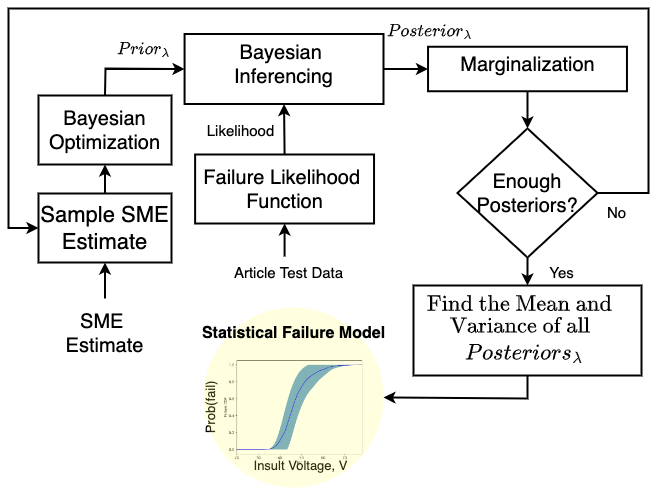}
  \caption{Bayesian Hierarchical Model Development using Limited Experimental Data. For the Hierarchy see fig. \ref{fig:failurecdfexample}. This process is repeated using a different sample from SME estimates, finding the mean and variance of the posterior$_{\boldsymbol{\lambda}}$, which is the statistical failure model.}
  \label{fig:FullModelStructure}
\end{figure}

\begin{figure}[htbp]
  \includegraphics[width=0.5\textwidth]{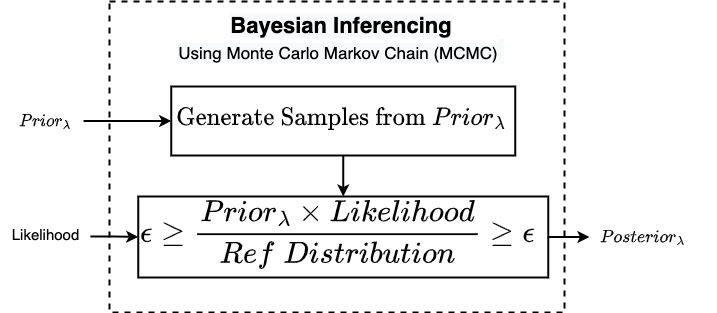}
  \caption{This is a more detailed version of the block "Bayesian Inferencing" found in fig. \ref{fig:FullModelStructure}.}
  \label{fig:BayesInfDiag}
\end{figure}

In this context, the population refers to the entirety of devices that exhibit sufficient similarity with respect to their response to an E1 HEMP-- as determined by a SME. Limited test observations are available for subsets of device Classes, specific Brands and Models. The application of hierarchy to Bayesian modeling uses related data to model the strength of the dependency between different groups based on Class, Brand, and Model.

\begin{figure}[htbp]
\centering
  \includegraphics[width=0.5\textwidth]{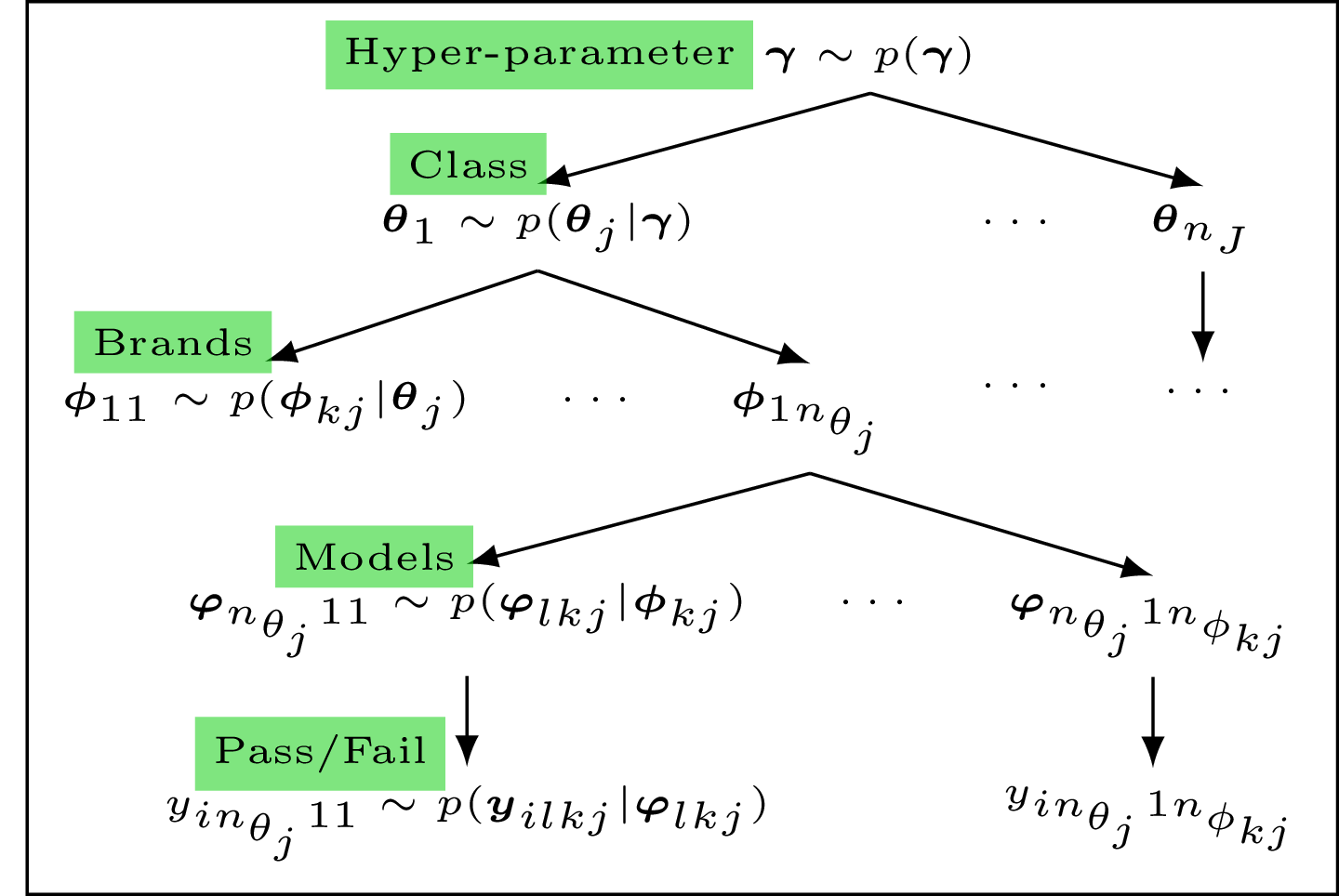}
\caption{Showing Hierarchy of Probability Relationships in the Bayesian Hierarchical Failure Model}
\label{fig:failurecdfexample}
\end{figure}

Our hierarchy identifies $n_J$ groups of device Class, with $n_{\boldsymbol{\theta}_j}$ device Brands in each type, and with $n_{\boldsymbol{\phi}_{kj}}$ device Models in each Brand (see Fig.\ref{fig:failurecdfexample}).
Hyperparameters parameterize the prior$_{\boldsymbol{\lambda}}$ distribution in a Bayesian hierarchical model, with the definition shown in Table.\ref{Tab:Tcr}. The choice of hyperparameters should be based on prior$_{\boldsymbol{\lambda}}$knowledge, if available, or on empirical evidence from previous studies. In Fig.\ref{fig:failurecdfexample}, $\boldsymbol{\gamma}$ is a set of hyperparameters. Choosing the hyperparameters which define the prior$_{\boldsymbol{\lambda}}$distribution $p(\boldsymbol{\gamma})$ enables us to simulate the probability of passing or failure of the device. This is also used to generate the posterior$_{\boldsymbol{\lambda}}$ from the device test data and SME estimate. Remaining parameters are the model parameters. We use $\sim$ to denote samples drawn from the corresponding distribution.

The posterior$_{\boldsymbol{\lambda}}$ over the parameter set,  $\boldsymbol{\lambda}:=(\boldsymbol{\gamma},\boldsymbol{\theta}_j,\boldsymbol{\phi}_{kj},\boldsymbol{\varphi}_{lkj})$, can be written using the Baye's formula \cite{18}.

The Baye's formula is implemented using Markov Chain Monte Carlo (MCMC) \cite{10} which is used for the Bayesian inference of the model parameter set  $\boldsymbol{\lambda}$. MCMC generates samples from the posterior$_{\boldsymbol{\lambda}}$ distribution over  $\boldsymbol{\lambda}$.  Samples from the posterior$_{\boldsymbol{\lambda}}$ are used to generate the statistical failure model (CDF) conditional on the applied voltage. The MCMC generates a sequence of samples which create the posterior$_{\boldsymbol{\lambda}}$ distribution as its equilibrium distribution. 

\begin{figure}[htbp]
\centerline{\includegraphics[width=0.45\textwidth]{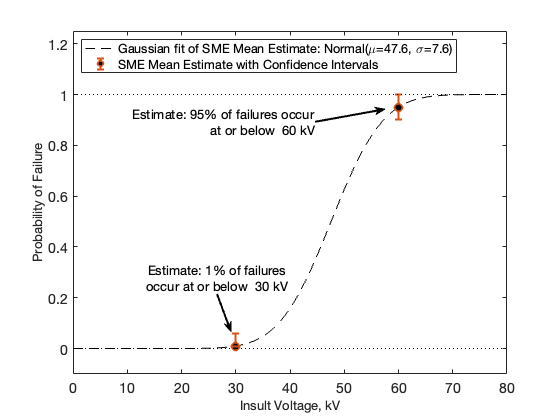}}
\caption{SME Estimates Shown In Terms of Assumed Gaussian Characterization for Simpler Observation. These are Developed by Fitting Subject Matter Expert Estimates to a Gaussian CDF. Estimated Confidence Intervals are also shown and discussed in Section VI.D. Only estimated points and CIs are used in BO to make the Prior$_{\boldsymbol{\lambda}}$.}
\label{fig:SME}
\end{figure}

The prior$_{\boldsymbol{\lambda}}$distribution over  $\boldsymbol{\lambda}$ should be carefully chosen. For  expensive components where limited testing is conducted, the statistical failure model (CDF) is largely driven by the prior$_{\boldsymbol{\lambda}}$selection. Therefore, input from one or more subject matter experts (SME) plays an important role in the prior's selection. We assume that the SME input comes in the form of a mean estimates of the CDF of component failure. An example of such SME estimate with zero variance is shown in Fig. \ref{fig:SME}., which has been defined by two points and an assumed Gaussian fit. As we will see, the SME estimate will be bounded by SME defined confidence intervals.

\color{black}
\subsection{Development of Prior$_{\boldsymbol{\lambda}}$ Distributions of Bayesian Hierarchical Model Parameters}
A prior$_{\boldsymbol{\lambda}}$ distribution plays a crucial role in Bayesian inference, especially when the available data is sparse. A prior$_{\boldsymbol{\lambda}}$ distribution provides an initial belief about the parameters of the model before observing the data. When the data is limited or insufficient, a well-chosen prior$_{\boldsymbol{\lambda}}$ can help regularize the model and prevent over-fitting. In the absence of prior$_{\boldsymbol{\lambda}}$ information, a flat or uninformative prior$_{\boldsymbol{\lambda}}$ distribution can be used, which assigns equal probability density to all values of the parameter. However, in many cases,prior$_{\boldsymbol{\lambda}}$ knowledge or beliefs about the parameter values are available, and incorporating this information can improve the accuracy of the inference.

We have two cases for selecting prior$_{\boldsymbol{\lambda}}$ distributions. In the first case, we have devices with no available test data or any prior$_{\boldsymbol{\lambda}}$ information about its failure characteristics. In the second case we have some information in the form of an SME estimate, as we discussed in Fig. \ref{fig:SME}.  The SME CIs are given as a mean value of a Gaussian distribution. However, these CI may not extend above 1 or below 0. Additionally the distributions themselves should not overlap to preserve the constraint that any CDF that are generated and which pass through these CIs must be monotonically increasing. Therefore Gaussian PDF representing both upper and lower CIs is defined with $\sigma$ to be equal to $1/1.96$ times the 5\% of the mean values. The upper PDF is then truncated at 1.0 above its mean and 0.5 below its mean. The lower PDF is truncated at 0.5 above its mean and 0.0 below its mean. The Gaussian assumption of the SME estimates do not imply a Gaussian posterior results from this process.  

\begin{figure}[htbp]
\includegraphics[width=0.45\textwidth]{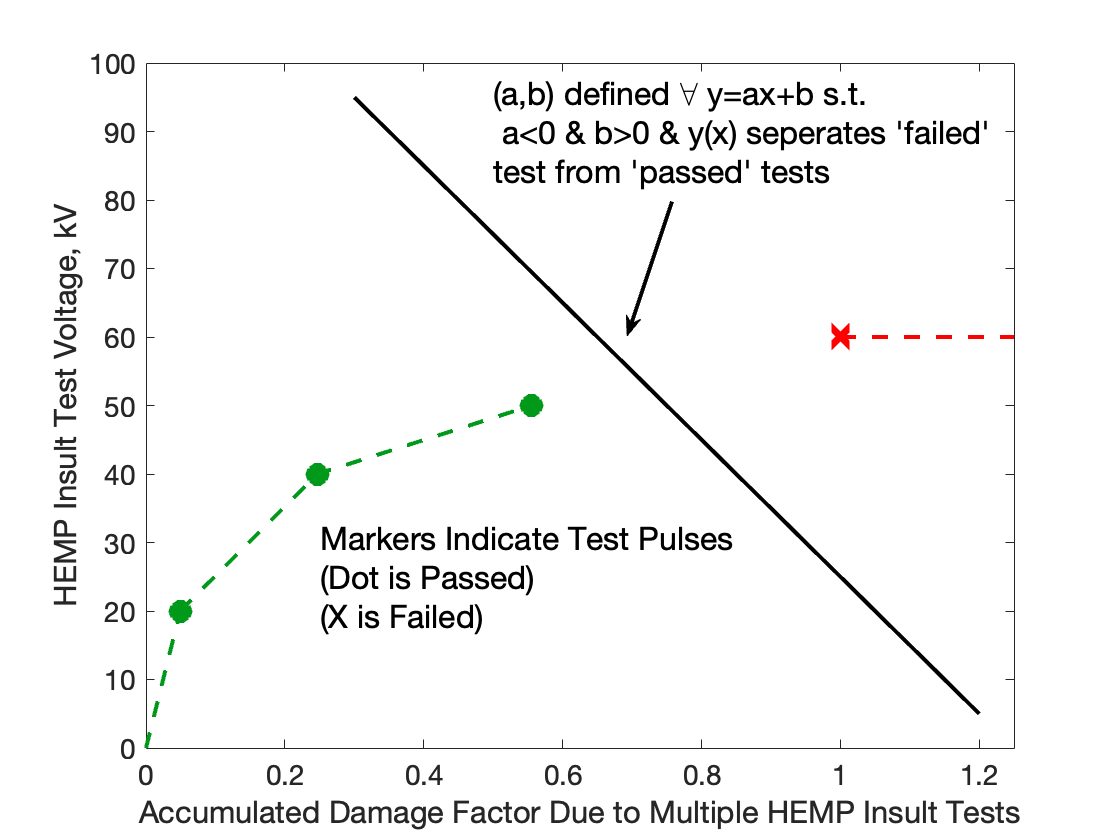}
\caption{ Solid line is one draw from the Bayesian Hierarchical Model which must separate failed test data (red X) from passed test data (green circles).}
\label{fig:expdataline}
\end{figure}

Normally the hyper-parameter selection is done using Bayesian Optimization as discussed later in Subsection \ref{ssec:BO}.
In absence of an SME estimated failure CDF, the hyper-parmater selections will be randomized within a constraint region and used to generate the prior. The prior$_{\boldsymbol{\lambda}}$ is represented by all the possible lines that exist which intersect the pass and fail regions (see Fig. \ref{fig:SME}) and which negative slope and a positive y-axis intercept. We note that the prior$_{\boldsymbol{\lambda}}$ for the model, should be such that the resulting device model parameters $\boldsymbol{\varphi}_{lkj}$ generate all lines that separates the pass and fail region as shown in Fig. \ref{fig:expdataline}. Our problem structure dictates that this prior$_{\boldsymbol{\lambda}}$ line will always have a negative slope with a positive x-axis and y-axis intercepts.  Although this process is laborious, owing to a large parameter tuning space of the prior$_{\boldsymbol{\lambda}}$ distribution parameters over  $\boldsymbol{\lambda}$, only a subset of them are highly sensitive. If SME inputs are available, the problem of prior$_{\boldsymbol{\lambda}}$ selection can be cast into an optimization problem, which minimized the distance ($L_{CE}$) per Eq. \ref{eq:cf} between the prior$_{\boldsymbol{\lambda}}$ and the SME estimate. 
We assume the SME inputs are available as a CDF , as shown in Fig. \ref{fig:SME}. We made a judgement to fit these data to a Gaussian CDF. However there is no precedent precluding a fit to other distribution types.

\begin{table}[t]
\centering
\caption{PDF for the Hierarchical Model, $\boldsymbol{\lambda}$}
\vspace{2mm}
\resizebox{\columnwidth}{!}{%
\begin{tabular}{ l | l |l}
    \textbf{Parameter $\boldsymbol{\lambda}$}& \textbf{Model parameter distribution} & \textbf{Meaning} \\
    \hline
    \hline
    $\alpha_i$ & $\text{TND}(\mu=0,\sigma_i,\alpha_{min} ,\alpha_{max})$& TND := Truncated Normal Distribution \\
    $\beta_i$ & $\text{TND}(\mu=0,\sigma_i,\beta_{min} ,\beta_{max})$&  \\
    \hline
    \hline
    $\mu_a(\alpha_i)$ & $\text{Uniform}(\alpha_1+\alpha_2, \alpha_1)$ & Model parameters\\
    $r_a(\alpha_i)$ & $\text{Uniform}(\alpha_3, \alpha_3+\alpha_4)$ & Model parameters\\
    $\mu_b(\beta_i)$ & $\text{Uniform}(\beta_1, \beta_1+\beta_2)$ & Model parameters\\
    $r_b(\beta_i)$ & $\text{Uniform}(\beta_3, \beta_3+\beta_4)$ & Model parameters\\
    $a_0$ & $\text{Uniform}(\mu_a-(r_a/2), \mu_a+(r_a/2))$& (Insult Voltage) $/$ (Cum. Insult Voltage)\\
    $b_0$ & $\text{Uniform}(\mu_b-(r_b/2), \mu_b+(r_b/2))$ & Insult Voltage
\end{tabular}%
}
  \label{Tab:Tcr}
\end{table}
\color{black}

Model parameters, $\boldsymbol{\lambda}$ are selected with distributions as shown in Table \ref{Tab:Tcr} and are used to develop the optimal prior, which minimizes the distance between the statistical failure model CDF and the SME estimates. Below, we outline the optimization technique we develop to generate the optimal prior$_{\boldsymbol{\lambda}}$ using SME estimates.

The SME estimates are a set of n points $\boldsymbol{\mathcal{X}}_{SME}:=(\boldsymbol{x}_1,\boldsymbol{x}_2,...,\boldsymbol{x}_n)$. Each of the $\boldsymbol{x}_i$ is a pair $(v_i,y^{SME}_i)$, where $v_i$ is the applied voltage to the device under test and $y^{SME}_i$ is the pass or fail outcome. This estimate also include confidence intervals. As an example, a 2-point SME estimate is used as an input to the Bayesian Optimization as shown in Fig. \ref{fig:FullModelStructure}. The Bayesian hierarchical model generates the statistical failure model CDF using the optimal prior$_{\boldsymbol{\lambda}}$ over $\boldsymbol{\lambda}$, prior$_{\lambda}$. 

\subsection{Bayesian Optimization}\label{ssec:BO}
When the prior$_{\boldsymbol{\lambda}}$ is generated, the CDF will not perfectly align with the SME estimates, nor will it necessarily  encompass the entire x-axis domain specified by the SME estimates. In these cases interpolation-extrapolation is used to align the x-axis values (insult voltages) of the SME estimates with the prior. We refer to the new aligned CDF values from the previous as $(y^{\boldsymbol{\lambda}}_1,y^{\boldsymbol{\lambda}}_2,...,y^{\boldsymbol{\lambda}}_n)$. We use optimization to minimize the distance function $L_{CE}$\cite{11} which is the cross-entropy function, where
\begin{align}
    L_{CE} = -\sum_{i=1}^n  y^{SME}_i\text{log}(y^{\boldsymbol{\lambda}}_i),   \label{eq:cf}
\end{align}
We use Bayesian-Optimization (BO) \cite{9} to minimize $L_{CE}$. BO is a powerful technique to optimize black-box functions that are computationally expensive to evaluate.  For a choice of decision variable, $\boldsymbol{\gamma}$, the statistical failure model gives a CDF of component failure.  

\begin{figure}[htbp]
\includegraphics[width=0.45\textwidth]{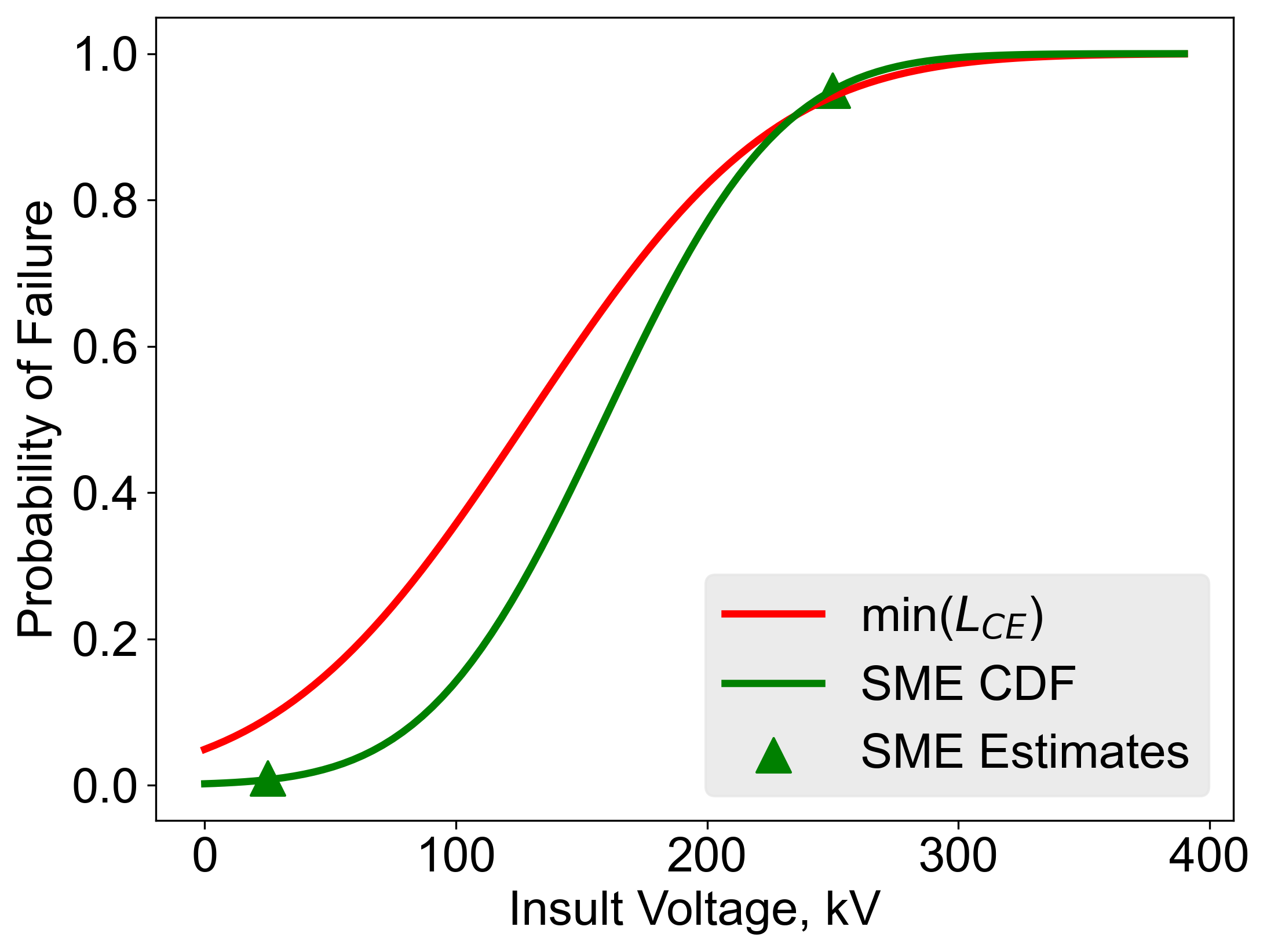}
\caption{Generating $L_{CE}$ Optimized Prior (prior$_{\boldsymbol{\lambda}}$) by passing SME estimates through Bayesian Optimization. This process is repeated by selecting additional samples within the SME Confidence Interval.}
\label{fig:tripcoildatasmeprior}
\end{figure}

\begin{figure}[htbp]
\includegraphics[width=0.45\textwidth]{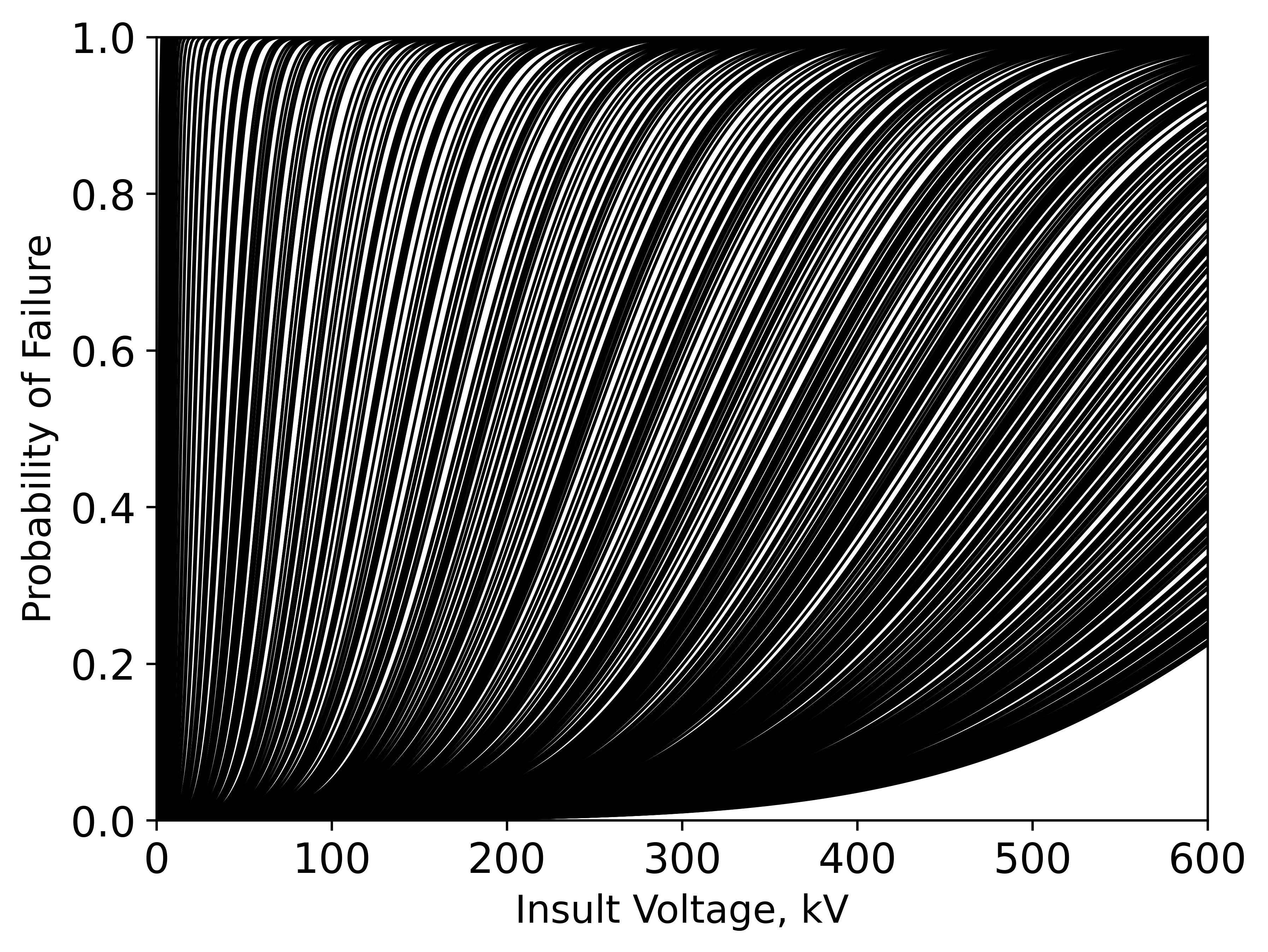}
\caption{Generalization of priors without SME estimates. All CDFs would result in covering the whole region, defined over all distribution.}
\label{fig:tripcoilcdf}
\end{figure}

In Fig.\ref{fig:tripcoildatasmeprior} we show the effect of $\boldsymbol{\gamma}$, the hyperparameter optimization to bring the prior$_{\boldsymbol{\lambda}}$ CDF closer to that of the SME estimates. If the SME estimates are not fused and $\boldsymbol{\gamma}$ is not optimized, then an infinite number of equally probableprior$_{\boldsymbol{\lambda}}$ CDFs will exist as shown in Fig.\ref{fig:tripcoilcdf}. 

\color{black}
\subsection{Method of Bayesian Failure Model (CDF) Development}
The statistical failure model of a device takes the form of a CDF, conditional on the magnitude of the voltage insult.
There are three steps to generate this CDF. First, we generate the prior$_{\boldsymbol{\lambda}}$ CDF as discussed in Section V-A. 
Second, we define the likelihood distribution of the statistical failure model: Given a realization of $\boldsymbol{\lambda}$ (which includes a specific insult voltage), how likely is the device to pass or fail, at that particular insult voltage?
Third, we perform MCMC, resulting in a joint posterior$_{\boldsymbol{\lambda}}$ distribution over all the model parameters, $\boldsymbol{\lambda}$ (Table I.), then marginalize the voltage variable $b_0$ over all of the tested devices. The resulting posterior$_{\boldsymbol{\lambda}}$ voltage samples are then combined to generate a histogram and a CDF is fit to it. This CDF is the statistical failure model. Traditional MCMC techniques such as Metropolis-Hastings requires a considerable amount of tuning. In contrast, SA-MCMC \cite{13} automatically adapts within its parametric family to best approximate the target distribution, making it more suitable for our application. We use the SA-MCMC implementation from the Python package \texttt{numpyro} \cite{14}. 

\section{Discussion on Statistical Failure Model Error}
While developing the statistical failure model we encounter various sources of error. We consider four causes: finite test measurements, error in SME prescriptions, Bayesian Inferencing computation, and computation in Bayesian Optimization.

\subsection{Error Due to Bayesian Inferencing Computation}  
Due to finite number of samples used in Bayesian Inferencing computation, we have errors in the representation of the prior$_{\boldsymbol{\lambda}}$ and posterior$_{\boldsymbol{\lambda}}$ CDF. We represent this error with a confidence interval around the posterior$_{\boldsymbol{\lambda}}$ CDF. This is done using the Dvoretzky-Kiefer-Wolfowitz inequality \cite{19}. To have the posterior$_{\boldsymbol{\lambda}}$ CDF within $\epsilon$ of the true CDF (defined by an infinite samples) with confidence $1-\alpha$, we choose the sample size $n$ using
 $$n\geq \Big(\frac{1}{2\epsilon^2}\Big)\text{ln}\Big(\frac{2}{\alpha}\Big).$$
For example, for the posterior$_{\boldsymbol{\lambda}}$ CDF to be within $0.01$ of the  true CDF with  $95\%$ confidence, we choose $n\geq18445$.  In our simulations we have selected the sample size to be $20000$ for the MCMC.

\subsection{Error Due to Bayesian Optimization Computation} 
Among the four error modalities identified in this section, the error caused by the Bayesian Optimization step is the most difficult to quantify. The Bayesian Optimization searches model parameters to  minimize the distance between the SME estimates and the prior$_{\boldsymbol{\lambda}}$ CDF which is the output of the Bayesian Optimization. The prior$_{\boldsymbol{\lambda}}$ CDF is the outcome of sampling in a hierarchy (see Fig.\ref{fig:failurecdfexample}), due to this, it is difficult to find the feasible set of the parameters. In Bayesian optimization we start with a boundary over the search space, where we are interested in finding the optimal parameters. 

Searching in the high dimensional space our parameters requires carefully choosing the search bounds for the optimization to complete within appropriate time. Searching the space requires sampling from distributions as shown in Table I.  

In one complete Bayesian Optimization search, we encounter a considerable number of infeasible candidate parameters. These are discarded, and the feasible parameters are used for proposing the next search location for the optima, based on an acquisition function. The assumption on the search bounds and finite number of search iterations leads to an inherent error that is added to the prior$_{\boldsymbol{\lambda}}$ CDF, which also gets reflected in the posterior$_{\boldsymbol{\lambda}}$ CDF. The error is then tuned to be negligible by carefully shrinking the search boundaries and dynamically selecting the optimization objective weights assigned to each of the individual SME estimates.

\subsection{Finite Testing Error in Failure CDF} 
Finite testing error is introduced when a finite number of test articles is used to determine the statistical failure model CDF. When dealing with empirical data that are financially expensive to obtain, the limited size of the test measurements can significantly impact the reliability of the posterior. To estimate the error in the statistical failure model CDF due to this finite testing error, we make the following assumptions. We assume no uncertainty in the SME estimates, negligible error in the Bayesian inference computation, and negligible contribution in the error from Bayesian optimization. 

\begin{figure}[htbp]
\centering
\includegraphics[width=0.45\textwidth]{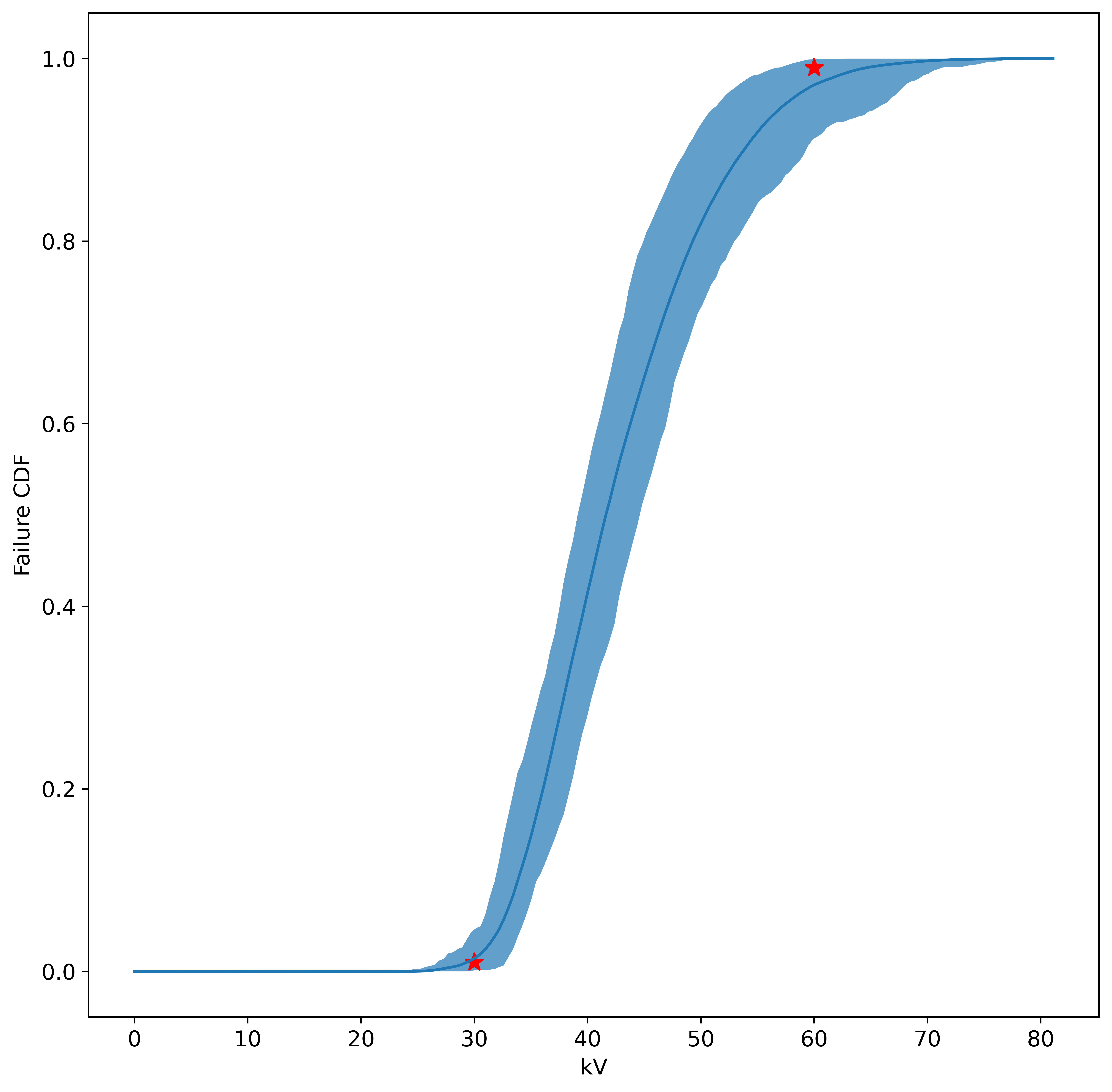}
\caption{ 
Error due to finite number of test articles in the postrior CDF.
This is the posterior CDF for 95\% CI. Red points are SME mean estimates.}
\label{fig:tripcoilfinitesample}
\end{figure}

Unlike traditional non-probabilistic and heuristic methods to generate a statistical failure model CDF, the Bayesian hierarchical  model has the advantage of providing interpretable failure CDFs using a small set of test measurements. However, a finite number of test articles leads to uncertainty in the statistical failure model CDF. We quantify this uncertainty by using different initial conditions in the MCMC chain, which generate the posterior$_{\boldsymbol{\lambda}}$ CDFs. The confidence bound of the statistical failure model CDF is calculated by combining these posterior$_{\boldsymbol{\lambda}}$ statistical failure model CDF as shown in Fig.\ref{fig:tripcoilfinitesample}. 

Although we have discussed as distinctly different forms of error, they are not directly additive.

\subsection{Error due to SME Confidence Interval Estimates} 
The SME estimates in this example has two mean values, one at the upper and one at the lower insult voltage, along with the $\pm$ error confidence intervals for each mean value.  The mean estimate of the SME is given by $(30 \ \text{kV},0.01), (60 \ \text{kV},0.99)$, (see Fig.\ref{fig:tripcoildatasmeprior}). This is interpreted as a $1\%$ chance of failure at or below 30 kV and a $99\%$ chance of failure at or below 60 kV.
The SME also defines the uncertainty of his estimates. In this case the SME has chosen a $95\%$ confidence interval (CI) which is $\pm 5\%$ of the mean estimates. This confidence interval is important because it effects the error in the prior$_{\boldsymbol{\lambda}}$.

\noindent \textit{Use of CI in Bayesian Optimization:} The BO uses the SME estimates and gives A prior$_{\boldsymbol{\lambda}}$ CDF that is close to those points. These two points are sampled individually from the uniform distribution through the extent of the confidence boundaries of each SME points. After iteratively doing this over N number of samples, BO calculates a family of hyper-parameters  that define a family of prior$_{\boldsymbol{\lambda}}$ CDFs. We can now define the mean and confidence interval of the prior$_{\boldsymbol{\lambda}}$ CDFs. This confidence is the error due to the SME estimates. 

\subsection{Including New or Realted Test Data in CDF Development}
The Bayesian Hierarchical Failure Model shown in Fig. \ref{fig:failurecdfexample} is applicable for device classes that are similar in the device failure characteristics. Often, the E1 HEMP test data for individual classes are sparse, due to cost, resulting in few tests. When components are similar, we may assume they have similarities in failure characteristics, then combine the test data to build a failure model. The proposed hierarchical model can include different classes, brands, and models of components. When new test data becomes available, we start with a previously calculated posterior$_{\boldsymbol{\lambda}}$ distribution as our new prior$_{\boldsymbol{\lambda}}$and repeat the process to get an updated posterior$_{\boldsymbol{\lambda}}$ incorporating the new data set. An updated SME estimate is not needed since it is already embedded with the previous posterior. We marginalize the updated posterior$_{\boldsymbol{\lambda}}$ to get the updated statistical failure model.

\section{Application of the Bayesian Statistical Method}
A breaker trip coil is a solenoid that provides a mechanical trigger to a breaker in a switchard, causing it to open (trip).  It operates when a low current passes through its coil, initiated by the breaker's protective relay.  These relays are connected to cabling that can incur a coupled high-voltage pulse from an E1 HEMP.  The power grid's response due to the potential failure of a coil is the inability to remotely trip a breaker during normal or emergency conditions. The consequences to the grid of many potential trip coil failures across a wide area must be determined in context of the grid. Using the limited test data available, the objective is to generate a statistical failure model for that component class. 

\subsection{Test Data of a Breaker Trip Coil}
HEMP effects on trip coils were tested and insulation ruptures were presented in \cite{15}. A high-voltage pulser was used to replicate the induced voltage waveform from a HEMP. The pulser is used to test breaker trip coils with increasing pulse magnitudes ranging from 20 kV to 80 kV. The State-of-Health of each trip coil was measured via mechanical operation and impedance measurements before and after each insult to identify any damage or degradation to the trip coils. Component insulation rupture was used to designate failure for each insult by an EMP conductive pulser in a laboratory setting, as shown in Fig. \ref{fig:tripcoildata}.
Although coils indicated insulation rupture as discussed, physical operability of all coils remained intact.  Test data for 12 identical trip coils was used to make the Bayesian statistical failure model.

\subsection{Generating A prior$_{\boldsymbol{\lambda}}$ from SME Estimates}
A prior$_{\boldsymbol{\lambda}}$ CDF for the trip coils did not exist so failure estimated were obtained from an SME .
The SME estimates for the failure CDF of the trip coil are shown in Fig. \ref{fig:SME}. The SME provided two a priori failure estimates.  The first estimate denotes $1\%$ failures that occur at or below 25 kV. The second estimate denotes $95\%$ of failures occurring at or below 175 kV. The SME selected both the voltage and the failure values. The SME also defined the uncertainty of his estimates. In this case the SME has chosen a $95\%$ confidence interval (CI) which is $\pm 5\%$ of the estimates. This confidence interval is important because it effects the error in the prior$_{\boldsymbol{\lambda}}$. 
Using these SME estimates we minimize the distance function $L_{CE}$, by optimizing over the Bayesian Hierarchical Failure Model parameter's ($\boldsymbol{\lambda}$) prior$_{\boldsymbol{\lambda}}$ distribution, shown in Table \ref{Tab:Tcr}. The resulting prior$_{\boldsymbol{\lambda}}$ CDF was developed as shown in Fig.\ref{fig:tripcoildatasmeprior}.

We used the SME's contribution to inform the prior$_{\boldsymbol{\lambda}}$ CDF before conducting Bayesian inference which also incorporates the trip-coil test data.  The values used to describe the prior$_{\boldsymbol{\lambda}}$ distribution are the outputs of the Bayesian optimization and are shown in Table\ref{tab:params}. 
\begin{table}[htbp]
\caption{Mean values of TND parameters in Table I.}
\centering
\begin{tabular}{|l|c|c|c|}
\hline
\hline
 Variable & High & Low & Scale \\
\hline 
\hline 
$\alpha_1$ & 0.4163 & -16.4337 & 0.7799 \\
$\alpha_2$ & 0.9114 & -14.6490 & 0.3252 \\
$\alpha_3$ & -2.0$\mu_a$ & 0.5046 & 0.4067 \\
$\alpha_4$ & -2.0$\mu_a - \alpha_3$ & -0.9018 & 3.0522 \\
$\beta_1$ & 12.2278 & 0.3111 & 3.9211 \\
$\beta_2$ & 4.1683 & 0.9645 & 4.3033 \\
$\beta_3$ & 2.0$\mu_b$ & -0.9957 & 2.7465 \\
$\beta_4$ & 2.0$\mu_b-\beta_3$ & -0.6837 & 0.3458 \\
\hline
\end{tabular}
\label{tab:params}
\end{table}

\begin{figure}[htbp]
\includegraphics[width=0.45\textwidth]{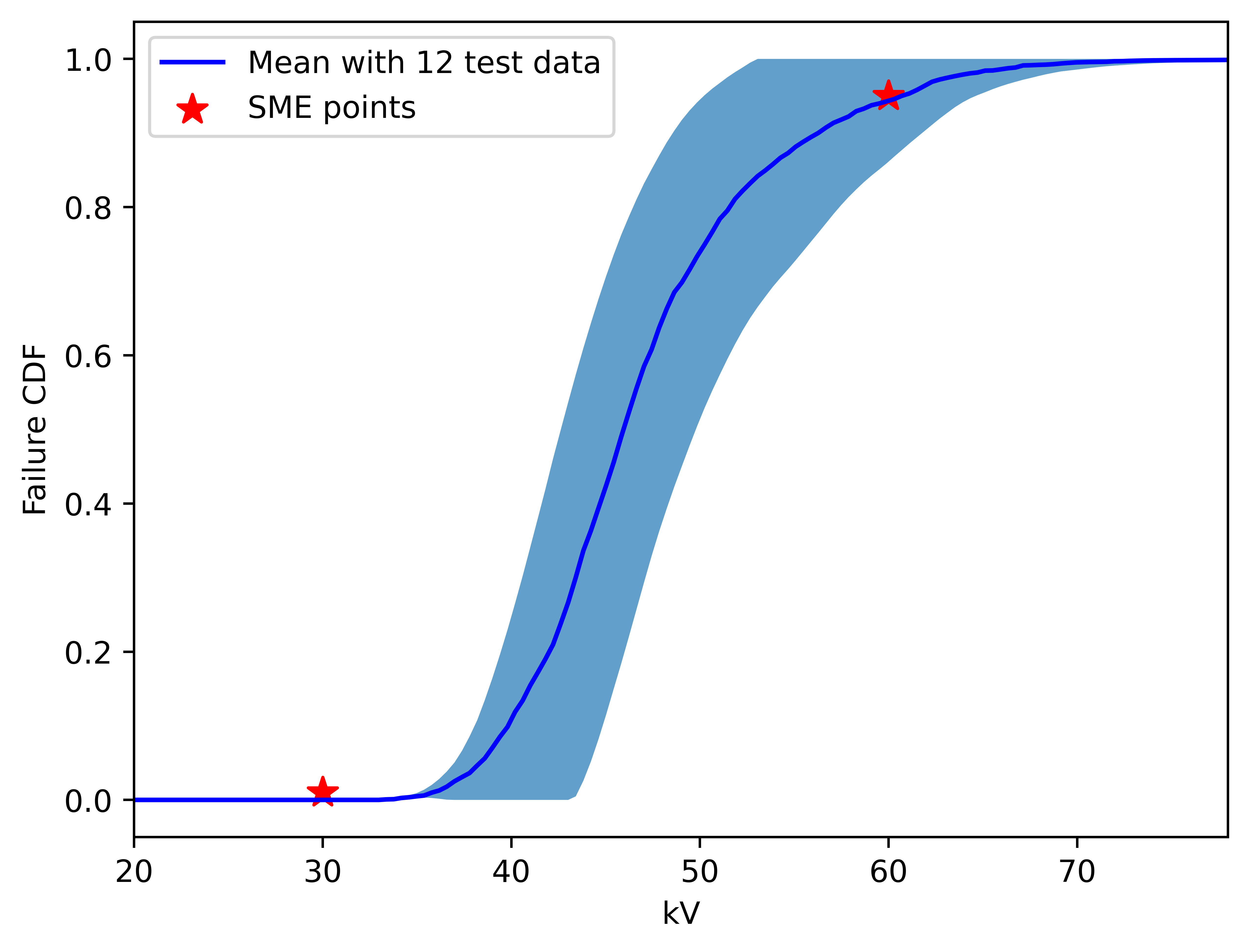}
\caption{Posterior device failure CDFs progressively representing the use of test results. These are generated by including all test data with the SME estimate. The shaded region indicates the 95\% CI, bounded within 0 to 1.}
\label{fig:tripcoilcdf4}
\end{figure}
\subsection{Integrating the Test Data with the prior${_{\boldsymbol{\lambda}}}$ to Generate Statistical Failure Model}

Given the SME truncated Gaussian distribution over CIs, we generate samples which are used in BO to calculate the samples of prior${_{\boldsymbol{\lambda}}}$. Each sample of  prior${_{\boldsymbol{\lambda}}}$ is then used in the Bayesian Hierarchical Model along with the test data to generate a candidate posterior failure CDF. For each sample of prior${_{\boldsymbol{\lambda}}}$ we are running the analysis shown in Fig 5 using all test data using the identical likelihood function, which returns a set of candidate posteriors. We identify the mean and standard deviation of the marginalized candidate posteriors, returning a final Statistical Failure Model, shown in Fig.\ref{fig:tripcoilcdf4}.  When sampling this model using MC methods, the use of variance may not be needed. However, the variance inherent to this model may be beneficial in determining the Uncertainty Quantification in the application that is using this statistical failure model.

We use the No-U-Turn Sampler \cite{16}, implemented in \texttt{numpyro}, instead of the SA-MCMC since it gives better sample variability. We discard the first 10000 samples from the posterior$_{\boldsymbol{\lambda}}$ distribution during the MCMC warm-up and then generate additional 20000 samples to represent the final posterior$_{\boldsymbol{\lambda}}$ distribution. This posterior$_{\boldsymbol{\lambda}}$ distribution is marginalized to establish the Statistical Failure Model CDF.

\section*{Conclusions}
Given the constraints of limited test data availability and the prohibitive costs associated with direct testing of E1 High-Altitude Electromagnetic Pulse (HEMP) effects on the electric power grid, the utilization of Bayesian priors presents a compelling approach for developing robust Statistical Component Failure Models. These models offer a feasible solution by facilitating simulations with minimal computational overhead, thereby enabling comprehensive assessments of the impact of E1 HEMP on large-scale power system interconnections.

As additional test information becomes accessible, it can be seamlessly integrated into existing Statistical Failure Models to enhance the accuracy of failure predictions. Moreover, Bayes' inference furnishes a potent framework for amalgamating diverse sources of information into a posterior Cumulative Distribution Function (CDF). Specifically, our investigation focuses on integrating Subject Matter Expert (SME) estimates and limited test data to refine our comprehension of device failure probabilities.

Expanding upon this approach, the incorporation of supplementary informative priors— such as SME estimates and related component posterior distributions— enables the application of convolution operations to merge their respective uncertainties. This convolution process serves to blend the encapsulated uncertainties within each distribution, thereby enriching the foundational beliefs regarding device failure characteristics. Consequently, this integrated approach fosters a deeper and more nuanced understanding of system vulnerabilities in the face of E1 HEMP events, paving the way for enhanced resilience strategies and risk mitigation measures within critical infrastructure.

\section*{Acknowledgment}

The authors are  thankful and grateful to  to the Sandia LDRD office for initiating this work under the EMP Grand Challenge and to Joe Blankenburg at DOE/CESER for sponsoring the completion of this paper.

\end{document}